\documentclass[12pt,usenatbib]{emulateapj}
\slugcomment{\small Manuscript edited: \today}
\usepackage{natbib}
\usepackage{epsfig}
\usepackage{color}

\newcommand{\kms}{~\rm{km\  s}^{-1}}

\begin{document}

\title{A close-pair analysis of damp mergers at intermediate redshifts}

\author{Richard C. Y. Chou\altaffilmark{1}}
\affil{Department of Astronomy and Astrophysics, University of Toronto, 50 St. George Street, Toronto, ON M5S 3H4}
\altaffiltext{1}{email: chou@astro.utoronto.ca}

\author{Carrie R. Bridge\altaffilmark{2}}
\affil{California Institute of Technology, 1200 East California Blvd. 91125 }
\altaffiltext{2}{email: bridge@astro.caltech.edu}

\author{Roberto G. Abraham\altaffilmark{3}}
\affil{Department of Astronomy and Astrophysics, University of Toronto, 50 St. George Street, Toronto, ON M5S 3H4\\}
\altaffiltext{3}{email: abraham@astro.utoronto.ca}

\begin{abstract}

We have studied the kinematics of $\sim 2800$ candidate close pair galaxies at $0.1<z<1.2$
identified from the Canada-France-Hawaii Telescope Legacy
Survey fields.  
Spectra of these systems were obtained using 
spectrometers on the 6.5m Magellan and 5m Hale telescopes.
These data
allow us to constrain the rate of dry mergers at intermediate redshifts and to test
the `hot halo' model for
quenching of star formation.
Using virial radii estimated from the correlation between dynamical and stellar masses
published by Leauthaud et al. (2011), we 
find that around 1/5 of our candidate pairs are likely to share a common
dark matter halo (our metric for close physical association). These pairs are divided into red-red, blue-red and blue-blue systems
using the rest-frame colors classification method introduced in Chou et al. (2011).
Galaxies classified as red in our sample have very low star-formation rates, but they need not be
totally quiescent, and hence we refer to them as `damp', rather than `dry', systems.
After correcting for known selection effects, the 
fraction of blue-blue pairs is significantly greater than that of red-red and blue-red pairs.
Red-red pairs are almost entirely absent from our sample, suggesting that damp mergers are
rare at $z\sim0.5$. Our data supports models with a short merging timescale ($<0.5$ Gyr)
in which star-formation is enhanced in the early phase of mergers, but
quenched in the late phase. Hot halo models may explain
this behaviour, but only if virial shocks that heat gas are inefficient until major mergers
are nearly complete.

\end{abstract}

\section{INTRODUCTION}
\label{intro}

The mass assembly history of the Universe is a major observable predicted by galaxy formation and evolution scenarios.  
Observational data have shown that galaxies in the local Universe generally fall into two categories: quiescent ellipticals, which dominate the massive galaxy population (M$_* \geq 10^{10.5} \rm{M} _{\odot}$), and blue star forming disc galaxies which occupy the lower mass regime \citep{baldry04, balogh04, kauffmann03}.
This color bimodality, which groups galaxies into the `blue cloud' and `red sequence',
 exists not only in the local Universe, but also out to redshift $z \sim 1$ \citep{bell04, willmer06},
and possibly up to $z = 2$ and beyond \citep{pozzetti10, conselice11}.   The origin and the evolution of this color bimodality is largely unknown. The
traditional view of red systems is that they are early-type galaxies that form monolithically at high redshift \citep{eggen62} though in modern theories of
galaxy evolution `nurture' plays as big a role as `nature', and evolution is driven by
several physical mechanisms, including
galaxy-galaxy interactions \citep{blumenthal84,dimatteo05,hopkins10,lambas12},  in-falling cold gas \citep{dekel06,dekel09,keres09,bournaud11},  and hot halo gas quenching  \citep{birnboim03,birnboim07,panuzzo11,gabor12}.

There is now a considerable body of evidence suggesting that the red sequence is populated by galaxies which
have somehow `migrated' there from the blue cloud \citep{martin07,hughes09,bundy10,shapiro10}.
Observations have shown that the total stellar mass of red massive galaxies (M$_* \geq 10^{11} \rm{M} _{\odot}$) has grown by 
a factor of two since redshift $z \sim 1$, while that of blue disc galaxies remains more or less the same \citep{bell04, faber07, abraham07}.  
Furthermore,  \citet{brammer11}  show that this mass growth trend can be traced up to $z \sim 2.2$.  
There is some evidence suggesting that the (visible wavelength) color transformation timescale for migration of a blue galaxy onto the red sequence is fast, $\sim 1$ Gyr \citep{bell04, blanton06}. 
Assuming the average gas to total mass ratio of late type galaxies is $\sim 15 \%$ \citep{rubin85} 
and a typical constant star formation rate is a few solar mass per year,  
it is difficult to consume all the gas content in blue galaxies within a timescale of 1 Gyr.  
In addition, simulations show that filaments in the intergalactic medium (IGM) can also channel new and fresh gas into
galactic halos, continuously feeding galaxies \citep{brooks09, keres09}. Thus, in order to move
a galaxy from the blue cloud onto the red sequence the star formation must be sharply cut off or quenched, rather than being stopped simply by a depleted supply of cold gas.  
Most mechanisms invoked for quenching star formation involve various forms of feedback which couples
star-formation in disks to hot gas in galactic halos \citep{gabor10}.

One well-studied mechanism for halting star-formation is the hot halo quenching model, which suggests that as gas falls into  galactic dark matter halos it is heated by virial shocks.  
If the mass of a dark matter halo exceeds $\sim 10^{12} \rm{M} _{\odot}$, the rate of gravitational heating by
shocks is greater than that of radiative cooling.
As a result, a hot virialized gas halo is formed near the virial radius. 
These hot halo shocks heat up the in-falling cold gas immediately,
 which prevents the gas from collapsing and forming new stars \citep{birnboim03, birnboim07}.  
 The hot halo quenching model alone cannot quench all 
star formation within a galaxy since some in-falling gas will still cool down and form stars in the center of the galaxy, but
this model provides a simple 
explanation for the mass-dependent bimodal galaxy distribution \citep{dekel06, cattaneo06, panuzzo11}.

Another interesting mechanism for halting star-formation in galaxies is major merger quenching.
In this picture, intense star formation triggered by the merging of two or more galaxies of roughly comparable stellar mass produces strong stellar winds that expel and/or heat the ISM
through shocks or feedback from supernovae \citep{cox06, ceverino09}.  
Stellar feedback from supernovae is probably not sufficient to quench
the star formation of the entire galaxy \citep{springel05a}, so
some additional energy source is probably required to supplement this quenching process.

Some semi-analytical models integrate these various ideas into a hybrid
picture in which a central super massive black hole (SMBH) is an additional energy supply.
In the hot halo quenching model, the in-falling cold gas could accrete on to the SMBH and trigger low luminosity AGN and radio emission, 
which is also known as the `radio' mode of the quenching process \citep{croton06, hopkins06a}. 
Similarly, major mergers can induce more violent material accretion and generate large amounts 
of energy feedback, which is known as the `quasar' mode of the quenching process. 
Both of these energetic feedback mechanisms involving the central SMBH could effectively expel or heat the surrounding gas disc and further quench the star formation activities \citep{hopkins06b}.

Irrespective of how quenching operates,
galaxy mergers are the central building blocks of the standard $\Lambda$-dominated cold dark matter model \citep[$\Lambda$-CDM;][]{cole08,neistein08}.
In this model major mergers transform 
disk galaxies into more massive spheroids \citep{toomre72,toomre77}.
However, it is difficult to study the full process of star formation and/or quenching mechanisms from single
snapshots of galaxy mergers. An alternative way forward is to 
study the evolution of large samples of mergers as a function of redshift, and
to connect these observations to models for triggering or quenching star-formation. 
One approach along these lines is to investigate the merger rate density evolution of  `wet' and `dry' mergers at different merging stages.
So-called wet mergers refer to galaxy mergers with intense star formation, while dry mergers are those with weak star formation.

Recent studies suggest
that the fraction of dry mergers is low at high redshifts ($z \geq 0.5$) 
but that dry mergers become important at  low redshifts ($z < 0.2$)
\citep{lin08,deravel09,depropris10,chou11}.
However, the reason for this change in the dominant mode
of merging, the mechanism by which color transformations occur, and the corresponding merging timescales
remain uncertain. A significant complication is that different procedures for defining
samples result in mergers with quite different properties. In essence,
there are  two main ways to define samples of mergers.
The first is based on morphological criteria, and focuses on structural disturbances and/or tidal features \citep{conselice00,conselice03,bridge07,jogee09,bridge10,chou11}.
Because mergers tend to show strong morphological disturbances at the first encounter and the last merging stage \citep{conselice06a,lotz08b},  morphological selection methods are biased toward selecting mergers at later merging stages.
The second approach is based on dynamical selection, which identifies mergers by limiting the projected physical separation 
and the velocity differences between two close pair members \citep{patton00,lin04,lin08,lopez10}.

Bridge et al. (2010) showed that when these different merger selections were normalized to the timescale 
in which they were sensitive to identifying mergers, the majority of studies agreed that the merger rate increases with redshift, 
however there remain some discrepancies in the absolute merger rate value at given redshift.  
Lotz et al (2010) suggest that differences in the range of mass ratios measured by different techniques and 
differing parent galaxy selection also likely contribute to the variation between studies.  

In this paper, our aim is to better constrain the number density of dry mergers at early merging stages, and to test the hot halo quenching model. 
Our analysis is based on imaging data from the Deep component of the  Canada-France Hawaii Telescope Legacy Survey (CFHTLS-Deep), 
and spectroscopic follow-up with the 6.5m Magellan and 5.1m Palomar Telescopes.  
Combining these datasets allowed us to define a sample of close kinematic pairs with redshifts between $z = 0.1 - 1.2$.
We investigate (1) the number of red-red, blue-red and blue-blue galaxy pairs as a function of stellar mass, 
and (2) the relationship between the visible-wavelength colors of close pairs and the galactic halo masses in which they reside.
In a companion paper, \citet{chou11}, we have constrained the dry merger number density at late merging stages,
and in this paper we seek to undertake a complementary analysis of systems in the early stages of merging.
By comparing the number density of dry mergers at different merging stages, we hope to 
clarify whether star formation is enhanced or quenched (or both, at different epochs) by mergers.

The paper is organized as follows:  \S2 gives the details of the sample selection,
 data reduction and the derivation of galaxy properties (stellar masses, color, spectroscopic redshifts, etc.)
Results are presented in \S3, followed by discussion in \S4. Our conclusions are presented in \S5.
Throughout this paper, we adopt a concordance cosmology with \hbox{$H_0$=70 km s$^{-1}$ Mpc$^{-1}$},
$\Omega_M=0.3$, and $\Omega_\Lambda=0.7$.

\section{OBSERVATIONS}
\label{obs}

\subsection{Photometry and galaxy Properties}

The optical photometry used in this paper primarily comes from two of the (CFHTLS) deep survey fields. Together these fields (denoted D1 and D4) cover an area of 2 square degrees. Another small fraction ($\sim 0.6 \%$)  of the sample comes from D2 field. The CFHTLS deep survey has high-quality broad-band photometry in five bands ($u^*$, $g'$, $r'$,$i'$,$z'$) and the depth of the survey ranges from 26.0 ($z'$) to 27.8 ($g'$).  The typical seeing for the final stacks is 0.7''-0.8'' in the $i'$-band.  
We utilized the imaging stacks from the Supernovae Legacy Survey \citep[SNLS]{sullivan06} for all our objects.  The source extraction and photometry were performed on each field using SExtractor \citep{ba96} in dual image mode.  The source detection was performed in the $i'$ filter ($i' \sim 26.3$), after applying a bad pixel mask, to avoid noisy or contaminated regions caused by spikes or halos of bright stars.  The total area masked is less than 10\% for each field. 
  
Galaxy properties such as stellar mass and star formation rate were derived by comparing the five broad band photometry to a set of template SEDs.
The best-fit SEDs were determined through a standard minimum $\chi ^2$ fitting between the template SEDs and the observed fluxes.  
The template SEDs were computed by the PEGASE-II galaxy evolution code \citep{fr97,lr02,leborgne04} and were convolved with the CFHT filters.  
A more detailed description of the SED  fitting analysis that was performed on this data set using the Z-Peg code \citep{bolzonella00,lr02} can be found in \citet{bridge10, chou11}.  
The stellar mass for each pair member was estimated using the Z-Peg code by integrating the total star formation history (SFH) of the best-fit model, 
up to the best-fit age and subtracting off mass loss from late stages of stellar evolution.  
The galaxy properties of close pairs were taken from the SNLS catalog.
An upper limit of photometric redshift $z < 1.2$ was applied for the sample presented in this paper because (a) there are a limited number of galaxies with robust spectroscopic redshifts $z>1.2$ to calibrate the photometric redhshifts, and (b) because the difference between the spectroscopic and photometric redshift becomes too large to derive accurate galaxy properties.

Although the galaxy properties of most of close pair galaxies are derived based on five broad band photometries and the best-fit photometric redshift,
we ran Z-Peg code again for close pair galaxies with confident spectroscopic redshift measurements (see Section \ref{sec:spec-z} for the measurement of spectroscopic redshifts) by replacing the best-fit photometric redshifts to confident spectroscopic ones to obtain precise galaxy properties. This indicates that the results presented in Figure 4 \& 5 are based on confident spectroscopic redshift measurements.
\begin{figure}
\begin{center}
\includegraphics[width=8cm,angle=0]{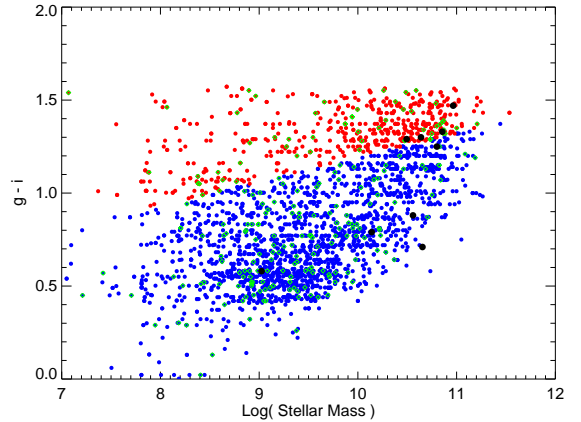}
\end{center}
\caption{Rest frame g-i color versus stellar mass for close pair galaxies in D1, D2 and D4 fields.  Red and blue dots represent the red and blue galaxies classified via the fiducial color as shown in Equation \ref{fiducial_color}.  Black dots are D1 galaxies with MIPS24 $\mu m$ detection.  Green diamonds represent close pairs which both galaxies are detected.  Note that red pairs also span a wide range in stellar mass, showing that there is no selection bias towards the bright red pairs. We artificially assign objects with 24 $\mu m$ detection as blue galaxies and excluded objects with photometric redshifts greater than 1.2. The sample end up with 607 red and 2143 blue galaxies. 
 }
\label{cmd}
\end{figure}

\subsection{Spectroscopy of the photometric sample}
\label{sec:sample}

This sample described in the previous section was used to define targets for a three-night spectroscopic program undertaken using the Inamori-Magellan Areal Camera and Spectrograph (IMACS) on the Magellan Baade 6.5-m telescope in October 2005 and September 2006. SCOPIC, a  {\bf{S}}pectroscopic study of {\bf{C}}lose {\bf{O}}ptical {\bf{P}}airs {\bf{I}}n {\bf{C}}FHTLS (PI Bridge), was aimed at identifying a large sample ($\sim$1500) of kinematic close galaxy pairs in order to further understand the galaxy merger rate and the connection between mergers, star-formation and AGN activity.   Close pair candidates were selected using the following criteria,  1) a projected separation between pair members less than 50 h$^{-1}$kpc, 2) an apparent $i^{'}$ magnitude difference $\le$ 1.5 mag, in order to select roughly equal mass or major mergers and 3) a photometric redshift difference between pair members of $\Delta  z < 0.15(1 + z)$.  This selection resulted in a sample of 2730 galaxies that resided in close pair (or triple system) of galaxies over 2 square degrees and were spectroscopically targeted.  

\begin{figure}
\begin{center}
\includegraphics[width=8cm,angle=0]{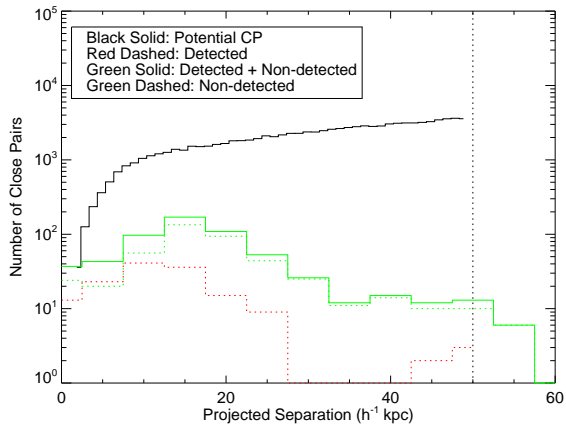}
\end{center}
\caption{ An analysis of the selection biases in our sample as a function of projected separation. The black curve represents the projected separation distribution of all the potential close pairs presented in D1 and D4 fields which satisfy the close pair selection criteria. The green solid and dashed curves represent the projected separation distribution of total and non-detected close pairs presented in this work; 
red dashed curve represents the projected separation distribution of detected close pairs (i.e., both pair members have confident redshift measurements, see Section \ref{sec:spec-z} for more details).
Note that there is a significant drop when the projected separation is greater than 20 $h^{-1}$kpc in the green and red curves, which indicates a selection bias for close pairs with smaller projected separations. 
}
\label{dis_SF}
\end{figure}

Figure \ref{dis_SF} shows the distribution of projected separations for the potential
close pairs in the fields chosen, along with our sampling of this distribution. 
The black solid curve represents the projected separation distribution for all the potential close pairs presented in D1 and D4 fields which satisfy the above mentioned selection criteria.
The green solid and dashed curves represent the projected separation distribution of total and non-detected close pairs, respectively. 
The red dashed curve represents the projected separation distribution for detected close pairs (both pair members have confident redshift measurements, see Section \ref{sec:spec-z} for more details). 
To increase the statistics, both green and red curves have a bin size of 5 $h^{-1}$\ kpc.
There is clearly a significant drop in our sampling of the parent population of potential close pairs when the projected separation is greater than 20 $h^{-1}$\ kpc, 
which indicates a selection bias for close pairs with smaller projected separations.
This selection bias is due to a higher selection priority of potential close pairs with projected separations smaller than 20 $h^{-1}$kpc.
Therefore, all the results presented in the present paper focus mainly on close pairs with projected separations smaller than 20 $h^{-1}$kpc.

Slits masks were designed to maximize the opportunity to get redshifts for both members of a close pair, by placing either a separate slit on each member or using a single slit that was tilted to cover both galaxies.  IMACS has one short (f/2) and one long (f/4) focal ratio camera.  The observations were conducted with the short camera
and the 8K $\times$ 8K Mosaic2 CCD detector, providing a field of view of 27$^{\prime}$.4 in diameter.  
We used the grating with a groove density of 200 lines/mm and the diffraction order $m$ of 1 to achieve the spectral coverage of 
5000 - 9000\,\AA \  with the central wavelength at 6600\,\AA. 
The slit width was set to $1^{''}$ and the wavelength dispersion was $\sim 10$\,\AA \ per slit or equivalent to a resolving power of 660.
The individual exposure times were 30 minutes with a total average integration time for each slit mask of 1.5 hours.
The data reduction was performed using the  Carnegie Observatories System for MultiObject Spectroscopy (COSMOS) software package,
developed by A. Oemler, K. Clardy, D. Kelson, and G. Walth and publicly available at http://www.ociw.edu/Code/cosmos.
We used COSMOS to align slit masks, correct flat fields, subtract sky background and perform flux calibrations for IMACS data.
The flux calibration stars were LTT1788 and gd71 for D1 and D4 fields, respectively. 
The spectral energy distribution (SED) of the standard stars were taken from \citet{hamuy92, hamuy94} and \citet{bohlin95}.
A 3$\sigma$ rejection threshold was used to remove cosmic rays when stacking sub-frames from the same slit mask.
1-D spectra were extracted using an IDL program we wrote for this purpose.

Although the vast majority of the data described in this paper comes from the Magellan Baade telescope via the SCOPIC survey, as a test of our methodology
we also
obtained spectra for a small sample of 20 (mainly red) close pairs in the CFHTLS D2 using the Palomar 200-inch telescope. 
The purpose of investigating this small sample was to verify that red galaxies are detectable using our techniques; we will occasionally 
refer to these Palomar observations when describing the effectiveness of our methodology of this paper, but to preserve sample
homogeneity we do not include any Palomar data in our
figures or analysis.
These pairs were observed using the Double Spectrograph (DBSP) on the Palomar 200-inch telescope in March 2010. 
DBSP has a blue and a red spectrograph that work simultaneously to provide a spectral coverage of 3000 - 7000\,\AA \ for the blue camera and 4700 to 10000\,\AA \ for the red camera, respectively.
The slit length of DBSP long-slit mode is 128$\farcs$ and the slit width varied from 1$\farcs0$ to $1\farcs5$ depending on weather conditions in order to achieve the highest S/N ratio.
We used a grating with a groove density of 1200 lines/mm that yields a wavelength dispersion of 1.38\,\AA \ and 2.1\,\AA \ for
 $1\farcs$0 and $1\farcs5$ slits, respectively at the diffraction order $m=1$.
The blazed wavelength of the red and blue spectrographs was set to 7100\,\AA \ and 4700\,\AA.
The individual exposure time was 10 minutes, with a total integration time of one hour per galaxy. 
The data reduction was done using a combination of standard IRAF routines and our own IDL code for extraction.

\subsection{Color classification}
\label{subsec_color}

Galaxies were classified as red or blue based on the stellar mass-dependent color selection criteria introduced in \citet{chou11}. In this
process galaxies are subdivided on the color-mass plane according to position relative to the following line:
\begin{equation}
\label{fiducial_color}
  (g' - i')_{\rm rest} = -0.0076+0.13\times M_{\rm stellar} - 0.1 
\end{equation}
The rest frame $g'-i'$ color is obtained from the 
If the rest frame $g' - i'$ color of a galaxy is greater than the fiducial color defined by the line, the galaxy is classified as red, and vice versa for blue galaxies. 
Figure \ref{cmd} shows the rest frame $g' - i'$ color versus the stellar mass of close pair galaxies in  the D1, D2 and D4 fields.
Red and blue dots represent red and blue galaxies, respectively. 
The black dots represent galaxies with the Spitzer Multiband Imaging Photometer for SIRTF
 (MIPS) 24 $\mu$m detection (down to a flux limit of 340 $\mu$Jy).
\citet{cowie08} report that at $z<1.5$ most red galaxies with a 24$\mu$m flux $>80 \mu \rm Jy$ fall into the blue cloud after the appropriate dust extinction corrections are applied. 
To account for the color change in dusty sources, we artificially assign the black dots with the rest frame $g'-i'$ color greater than the fiducial color to the blue cloud. 
Note that MIPS 24 $\mu$m observations do not include the D4 field, so the black dots on the Figure \ref{cmd} correspond to galaxies from the D1 field only.
However, there are only 10 objects in the D1 field with a MIPS 24 $\mu$m detection, which is less than 1\% of the sample in the D1 field.
It is reasonable to assume that the number of galaxies with MIPS 24 $\mu$m detection in the D4 field is similarly low and therefore this should not affect the overall result.
After the correction for 24 $\mu$m detections, the sample is comprised of 607 red galaxies and 2143 blue galaxies.
Green diamonds on Figure~1 represent close pairs where both members have spectroscopic redshift measurements.  

\subsection{Spectroscopic redshifts}
\label{sec:spec-z}
Redshifts were obtained from our spectra based on emission line identifications and cross-correlation against blue and red galaxy templates.
The emission line identification method is straightforward by recognizing multiple emission lines using their relative positions on the wavelength axis.
The redshift is determined by fitting a Gaussian function to the identified brightest emission line.
The corresponding redshift and the $1\sigma$ error can be derived from the central wavelength and the $1\sigma$ error of the fitting function.
For the majority of the continuum detected galaxies, emission lines are obvious and relatively easy to identify, however in a small number ($\sim 20\%$) of cases, emission lines are hard to identify or only the continuum can be seen (e.g., very red elliptical galaxies).
For those galaxies, we used the cross-correlation fitting method to determine the spectroscopic redshift by identifying absorption features.
We developed an IDL program based on the package `C\_CORRELATE' and adopted the photometric redshift as a prior guess.
To save computation time, we only considered the redshift range of $z_{pri} \pm 1.0$ (if $z_{pri} < 1$ then we start from $z=0$).
A blue and a red galaxy template generated from the stellar population synthesis code BC03 \citep{bc03} were used during cross-correlation fittings.
The templates were constructed using a model that convolves a single stellar population with solar metallicity and 
a exponentially declining star formation history with an e-folding timescale $\tau = 1$ Gyr.
The star formation cutoff time t$_{cut}$ was set to 20 Gyr.
We also applied the dust attenuation to templates using the default values in BC03 that the total effective $V$-band optical depth $ \tau _V $ equals to 1.0
and the fraction $\mu$ of the contribution from the ambient diffuse ISM equals to 0.3, respectively.
Finally, we took the SED with the age of 1.0 Gyr as the blue galaxy template, and the SED with an age of 5.0 Gyr as the red galaxy template.
Since the output SEDs from the BC03 do not include the emission lines, we added emission lines artificially to the blue template based on the 
emission line list used in the DEEP2 survey kindly provided by Renbin Yan (private communication).  The emission line widths in the DEEP2 line list were convolved to the IMACS spectral resolution before being added to the blue template.

\begin{table*}[bp]
\caption{Redshift confidence classes}
\begin{center}
{\tiny \begin{tabular}{lccl}
\hline
Class & Confidence & Number of Galaxies & Note \\
\hline
Failures: & & \\
\  \  \ 0 & None & 1367 &No detection.  \\
\  \  \ 1 & $< 50 \%$ & 270 & Only one emission/feature is detected. \\
    &        &            & The emission line can not be identified even  \\
    &        &            &  with the aid of photometric redshift. \\
Redshifts inferred from multiple features: &  &  & \\
\  \  \ 2 & $> 75 \%$ & 321 &  Reasonably secure. Redshift determined  \\
           &                     &         &  from two or more emission lines or features.  \\
\  \  \ 3 & $> 95 \%$ & 327 &  Secure. Redshift determined from multiple  \\
           &                     &         &  emission lines/features $+$ supporting continuum.  \\
\  \  \ 4 & Certain & 168 & Unquestionably correct.   \\
Single Line Redshifts: & &  &  \\
\  \  \ 9 &              &  297     & Obvious single line emission.  \\
           &              &              & The inferred spectroscopic redshift is consistent    \\
\  \  \    &              &              & with the photometric redshif.  \\
\hline
\end{tabular}
}
\end{center}
\label{confidence}

\end{table*}

We adopted the confidence level classification system introduced by \citet{abraham04} for the reliability of the spectroscopic redshifts (see Table \ref{confidence} for definitions of the confidence levels).
Class 1 corresponds to an an unreliable redshift, while class 4 corresponds to a high reliable redshift.
Single emission line objects require special consideration because in most cases a line is detected with certainty, but
the identification of the line is uncertain.
In such cases we used photometric redshifts as an auxiliary source of information, and
if the inferred spectroscopic redshift is within 1$\sigma$ error of the photometric redshift, we classify the object as having a
high confidence redshift (class 9), otherwise we assign it a low confidence level (class 1).

\begin{figure}
\begin{center}
\includegraphics[width=9 cm,angle=0]{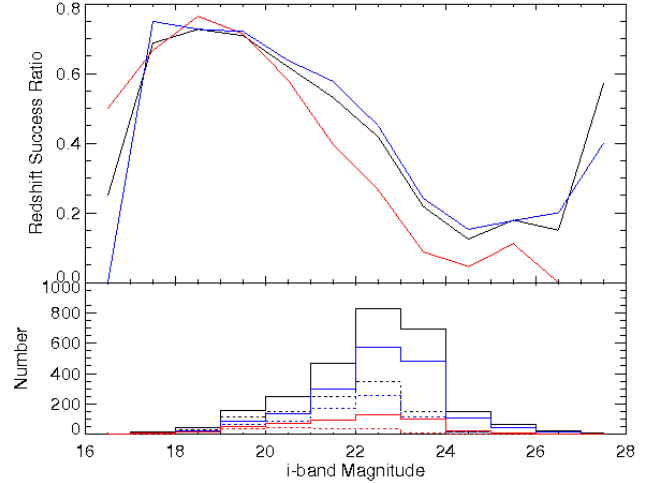}
\caption{(Upper panel) Successful redshift ratio versus $i^{'}$-band apparent magnitude.  The black solid line shows the successful ratio for total galaxies; blue and red solid lines shows the successful ratio of blue and red galaxies, respectively.  (Bottom panel) Histogram corresponding to the total, red and blue galaxies shown in black, red and blue, respectively.  Solid histograms represent the total number of galaxies in different magnitude bins, while dashed histograms represent the number of galaxies with  confidence level greater than 1.
}
\label{SF}
\end{center}
\end{figure}

Of 2750 galaxies targeted with slits, roughly half of them (1385) have redshift measurements.  
About $80\% $ (1115) of these redshifts have confidence levels greater than 1, which we refer to successfully detected ones.
As expected, the redshift success rate is directly related to the brightness of the galaxy, and Figure \ref{SF} shows our magnitude selection
function (the fraction of successful redshift measurements as a function of $i'$-band magnitude) for the total sample and for red and blue subsets.
For galaxies with an $i'$-band magnitude brighter than 20, the success rate for blue galaxies is more or less equal to that of red galaxies, 
while at fainter magnitudes ($i'>$20 mag) the rate is on average $\sim$20\% higher for blue galaxies.  This is understandable, as blue galaxies, having active star formation, often exhibit emission lines which are easily detected even when continuum measurements needed for absorption-line redshifts are impossible to obtain. The lower panel of Figure \ref{SF} shows the absolute number of galaxies in our sample as a series of histograms.
The solid lines represent the total number of galaxies, while dashed lines represent the total number of galaxies with confidence levels greater than unity.
The histograms show a fairly broad peak spanning $22<i'<24$~mag, and at this point the selection function
corresponds to $\sim$ 40\% successful detections
\footnote{The success rate is similar for the sample of 20 galaxies used to test our methodology with the Palomar 200-inch.
Roughly half of the Palomar galaxies were not detected due to bad weather conditions, although nine pairs were observed in good weather conditions.
We obtained six spectroscopically confirmed close pairs (12 galaxies) out of nine pairs (18 galaxies), 
which corresponds to a redshift success rate of $\simeq 70\%$.
Among the six spectroscopically confirmed close pairs, one was a blue-blue pair and the remainder were red-red pairs.
The members of these pairs have $i'$ band magnitudes brighter than 20.}.

In summary, 
we have obtained redshifts for 1385 galaxies in the range $0.1 < z < 1.2$, and 1115 of these ($\sim 80\%$) have high-confidence measurements.
Table \ref{confidence} shows the number of galaxies in each confidence level.

\section{RESULTS}
\label{results}

\begin{figure*}[htbp]
\begin{center}
\includegraphics[width=13cm,angle=0]{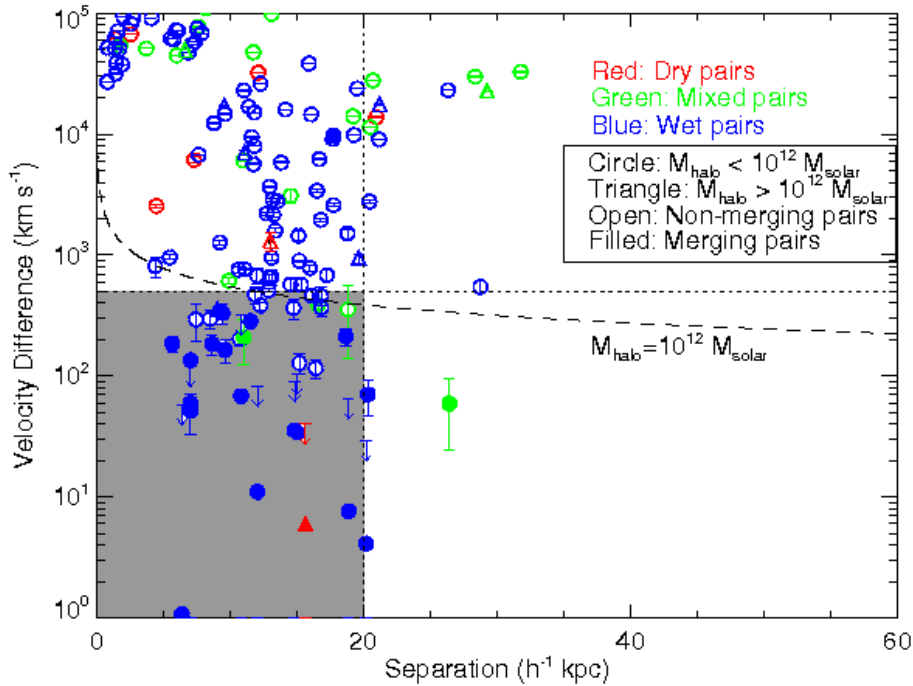}
\end{center}
\caption{ Diagram of separation versus velocity difference for close pair galaxies.  Red, green and blue colors represent red-red, blue-red and blue-blue pairs, respectively.  Circle and triangle symbols represent galaxies with their total halo masses smaller and greater than $10^{12} M_{\odot}$ (equivalent to a stellar mass of $10^{11} M_{\odot}$ using the stellar to total halo mass ratio of 0.1 taken from \citet{gonzalez07}). Filled symbols represent pairs with their separation smaller than the corresponding virial radii, which are classified as merging pairs. Two dotted lines indicate the separation of $20 h^{-1}$ kpc and the velocity difference of $500~\rm{km s}^{-1}$. The bottom left grey corner defined by the two dotted lines shows the conventional real merging pair selection area suggested by \citet{patton00}. The dashed lines represent the virial radius as a function of velocity difference of a system with a halo mass of $10^{12} M_{\odot}$. Before the selection of merging close pairs, there are 148 spectroscopic pair candidate (8 red-red pairs, 29 blue-red pairs and 111 blue-blue pairs on the plot).   Note that none of the close pairs presented on the diagram has MIPS24 $\mu$m detection (flux limit equals to 340 $\mu$Jy).   After the pair selection based on the virial radius assumption using the stellar to total halo mass ratio equals to 0.1, 21 out of 148 pairs are merging pairs. The 21 selected pairs is composed of one red-red pairs, two blue-red pairs and 18 blue-blue pairs. }
\label{sv-diagram_gon}
\end{figure*}

Figure~\ref{sv-diagram_gon} shows the
relative velocity versus physical separation for all close pairs in our sample
in which both members have high-confidence redshifts. 
The format of this figure resembles that of Figure~5 in \citet{patton00}, but with
a number of key differences, which we will now proceed to describe.

\begin{figure*}[htbp]
\begin{center}
\includegraphics[width=13cm,angle=0]{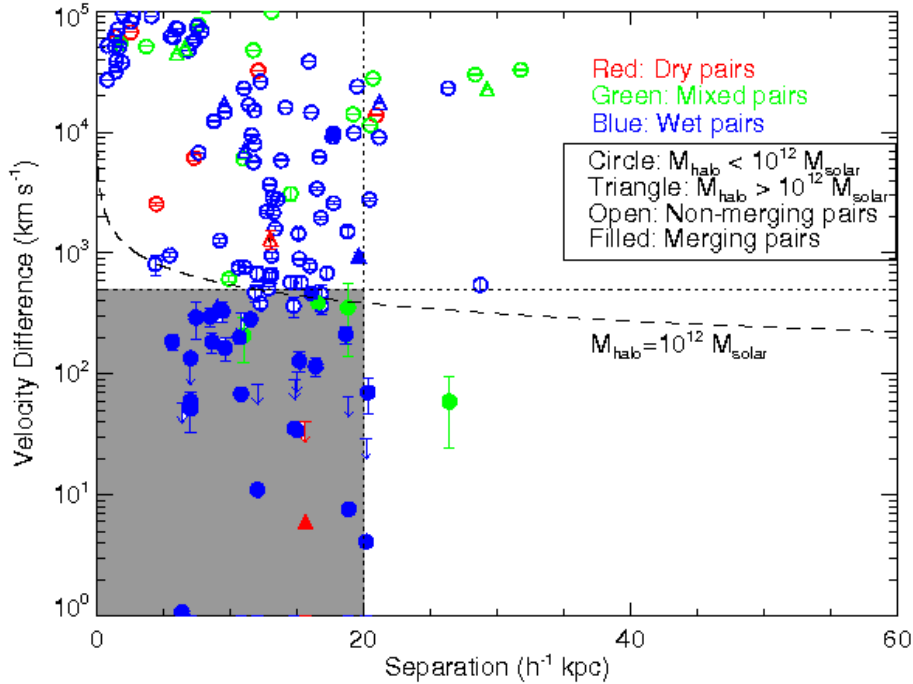}
\end{center}
\caption{ Diagram of separation versus velocity difference for close pair galaxies using different stellar to total halo mass ratios taken from \citet{leauthaud11}. This figure is presented in the same way as Figure \ref{sv-diagram_gon}.  It is clear that the number of filled symbols increases, and most of them are still located within the conventional merging pair selection area.  After the application of different stellar to halo mass ratios, 30 out of 148 are classified as merging pairs. One of them are red-red pairs, four of them are blue-red pairs and 25 of them are blue-blue pairs. Table \ref{pair_sum} summarizes the number of pairs selected by different criteria.}
\label{sv-diagram_new}
\end{figure*}

The central difference between our analysis and that of \citet{patton00} is that we have attempted to sub-divide pairs by the 
stellar populations of their member galaxies.
We distinguish between `red-red' pairs (systems in which both galaxies are quiescent), `blue-red' pairs (systems in which one galaxy is quiescent while the
other is star-forming) and `blue-blue' pairs (systems in which both galaxies are star-forming). Red, green and blue colors in Figure \ref{sv-diagram_gon} 
represent red-red, blue-red and blue-blue pairs, respectively. 
Since none of the galaxies presented on Figure \ref{sv-diagram_gon} and Figure \ref{sv-diagram_new} have Spitzer MIPS $24\mu$m detection, contamination of the quiescent red galaxy sample by very dusty star-forming systems is probably negligible.
There are 148 pairs (exclude the biased selected D2 samples) selected on the basis of photometric redshift where both members have confident spectroscopic redshift measurements, 
and we will show in a moment that only around one-fifth of these turn out to be physical pairs
once spectroscopic redshift and velocity information is incorporated.
Among the above-mentioned 148 candidate pairs, 8 are red-red pairs, 29 are blue-red pairs and 111 are blue-blue pairs.

In addition to segregating pairs by their dominant stellar populations, Figure~\ref{sv-diagram_gon} also subdivides pairs into two bins
on the basis of their estimated total ({\em i.e.} dynamical) masses. We adopt the simplistic approach of estimating the system's total mass by simply scaling each galaxy's stellar mass by a constant total mass-to-stellar mass ratio of 10:1 taken by \citet{gonzalez07} and then summing the masses (we will describe a somewhat more refined approach later in this paper). Subdivision into high-mass and low-mass bins occurs at a total mass of $10^{12} M_{\odot}$ because, as noted earlier, this is the mass threshold predicted by the hot halo model.
Circles and triangles in Figures~\ref{sv-diagram_gon} and 
\ref{sv-diagram_new} represent pairs with total halo masses less than and greater than $10^{12} M_{\odot}$. 

\citet{patton00} note that close pairs with separations less than $20 h^{-1}$ kpc and 
velocity differences less than $500~\kms$
generally show clear morphological signs of galaxy-galaxy interactions.
Our own visual inspection, described earlier, concurs with Patton et al.'s impression
and we therefore adopt these criteria as a useful starting point for our own more
detailed analysis.
Patton et al.'s limits are shown
as two dotted lines on Figure \ref{sv-diagram_gon},
and in the present paper we will treat
pairs inside these limits (the light grey region
at the lower-left corner of the figure) as probable merging pairs. 
Figure \ref{sv-diagram_gon} shows a gap at separations smaller than 5 $h^{-1}$ kpc and velocity differences $\Delta v < $10,000  km $s^{-1}$.
This gap finds its origin in an inevitable ambiguity that arises when distinguishing between close pairs
and on-going mergers at very small separations. The former were targeted by our investigation, but the latter
were not.  To verify our explanation for the gap in the figures, 
we inspected both the two-dimensional spectra and the $i'$-band images of all galaxies seen at small separations, 
and compared them with merging systems from the sample in Bridge et al. (2010). 
Objects located at the upper left corner ($\Delta v > $10,000  km~s$^{-1}$) of the figures correspond to physically unconnected pairs seen at small separations because of projection effects.
While these objects were generally both placed on a single slit, 
they exhibited no obvious tidal features on the $i'$-band images, 
so that the catalog construction process and subsequent morphological classification treated them as independent galaxies. 

In addition to the gap at small separations, Figure~\ref{sv-diagram_gon}  shows a paucity of close pairs with  separations greater than $20 h^{-1}$ kpc.
This finds its origin in the selection bias against systems with large separations described in Section \ref{sec:sample}.


If we adopt Patton et al.'s methodology, there are 31 probable merging pairs isolated in Figure~\ref{sv-diagram_gon}. It is interesting to note that
27 of these are blue-blue pairs, three are blue-red pairs and one of them is a red-red pair. 
This strongly suggests that red-red mergers are quite rare at $z\sim0.5$ (at least for close pairs with their separation smaller than $20 h^{-1}$ kpc), which is consistent
with the reported decline in `dry' merging activity at such redshifts reported
in Chou et al. (2011).
To our magnitude limit, blue
galaxies outnumber red galaxies by nearly a factor of four: the parent sample contains 607 red galaxies ($\sim 22\%$ of the total sample) and 2143 blue 
galaxies ($\sim 78\%$ of the total sample). The most conservative assumption is to assume that merging has no impact whatsoever
on the colours of galaxies and that galaxies of different colour are spatially uncorrelated, in which case the
colours of close pairs can simply be modelled as a binomial distribution with a 22\% probability that a galaxy is red. In this case 
we expect the fraction of red-red pairs to be $(0.22)^2$, 
corresponding to a $4.8\%$ red fraction, and therefore we expect to see 1.5 red-red pairs amongst our sample of 31 physical pairs. 
This is consistent with the single detection of a red-red merging pair.

However, a legitimate criticism of the analysis presented above is that
our definition for what constitutes a probable merging pair is non-unique. For example,
\citet{lin04, lin08} advocate a maximum separation of $50 h^{-1}$ kpc rather $20 h^{-1}$ kpc. If we extend our sample out
to this limit, there would be one more blue-red and two more blue-blue merging
pairs\footnote{It is worth recalling that the Milky Way is about 750 kpc from M31 with
a velocity difference of about $120\kms$ \citep{karachentsev04,devau91}. Even though our ultimate destiny seems to be to merge
with the Andromeda galaxy, the present configuration of the Local Group would not result in the Milky Way + M31
system being flagged as a merging pair by any of the criteria adopted here.}. 
Given the somewhat arbitrary nature of
the limits chosen, it is worthwhile considering whether or not a more physically motivated definition
for what constitutes a merging pair might not be preferable. 
Therefore, inspired by the idea of a halo occupation number \citep{linYT04,hopkins10},
we propose the following simple definition: \textbf{a merging pair is a system of two galaxies whose projected separation $r_p$ 
is equal to or less than two times the total system's virial radius.} 
In other words, the pair can be thought of as sharing a common halo if
$r_p < 2\, r_{\rm virial} $, where $r_{\rm virial} = G M/ V^2$, and
$G$ represents the gravitational constant, $M$ represents the total halo mass of the system and $V$ represents the 3D relative velocity of the two pair members.
Since the actual 3-D relative velocity is not available, 
we must multiply line-of-sight velocity differences by a factor of $\sqrt{3}$ as a statistical correction when computing $r_{virial}$ from redshifts
\footnote{The factor of $\sqrt{3}$ is obtained by assuming an isotropic
velocity difference distribution, so that the relationship between the pairwise
velocity $V$ its components $V_x, V_y$ and $V_z$ can be simply
expressed as  $\sqrt{3}$  times the line-of-sight velocity:
$V=\sqrt{V_x^2+V_y^2+V_z^2} = \sqrt{3 V_z^2} = \sqrt{3} V_z$.}.

For reference, 
the virial radius of a system with a total mass of $10^{12}~M_{\odot}$ and a velocity difference of 500 km s$^{-1}$ is $12~h^{-1}$ kpc.
The dashed curve in the figure maps variation of virial radius as a function of velocity difference 
for a system with the halo mass of $10^{12}~M_{\odot}$. In our approach any system with a total mass of $10^{12} M_{\odot}$ would be considered
a merging pair if it lies anywhere below this curve. We applied this methodology to systems of all
masses by assuming a certain total mass-to-stellar mass ratio, 
and in Figure~\ref{sv-diagram_gon} we flag merging pairs determined in this manner using filled symbols. 
Using this approach, 21 systems (around 65\% of the pairs in the boxed region at the lower-left corner of Figure~\ref{sv-diagram_gon}) are found to be merging. 
One of these is a red-red system and two pairs are blue-red systems, but 18/21 (86\%) of the pairs are  blue-blue systems.

\begin{table*}
\caption{Summary of three types of close pairs selected using different criteria.}
\begin{center}
\begin{tabular}{lccc}
\hline
 criteria & $r_p < 20 h^{-1}$kpc & $r_p < 2\, r_{virial}$  & $r_p < 2\, r_{virial}$ \\
              & $\Delta v < 500 \rm{km s}^{-1}$ & (\citet{gonzalez07}) & (\citet{leauthaud11}) \\
\hline              
 Red-red Pairs  &  1  &  1  &  1 \\
 Blue-red Pairs &  3  &  2  &  4  \\
 Blue-blue Pairs &27  &  18  &  25  \\
\hline
 Total  &  31  &  21  &  30 \\
\hline
\end{tabular} 
\end{center}
\label{pair_sum}
\end{table*}

A criticism that can be directed at Figure~3 is its heavy reliance on 
a uniform stellar-to-total mass ratio in order to infer virial radii.
As
described earlier, the total masses (and hence virial radii) used to construct Figure~\ref{sv-diagram_gon} have been computed using
the rather simplistic approach of assigning a total mass to a composite system by assuming a constant total mass-to-stellar mass ratio
of 10:1. This commonly-adopted \citep{conroy07,spitler08} total mass to stellar mass ratio is taken from \citet{gonzalez07}, who derived it from 
galaxy groups and clusters with total halo masses $\ge 10^{14} M_{\odot}$, which are about two orders of magnitude more massive
than the total masses of the systems in our sample.
Naturally, other measurements of the total mass-to-stellar mass ratio can be found in the literature
(for example \nocite{conroy07} Conroy et al. 2007, \nocite{giodini09} Giodini et al. 2009, \nocite{Leauthaud10} Leauthaud et al. 2010 ) but 
most such work focuses either on limited classes of objects (e.g. early type galaxies; 
\nocite{heymans06} Heymans et al. 2006) 
or on composite systems with masses higher than $10^{13} M_{\odot}$ \citep{mandelbaum06}. 

Fortunately, statistical analyses of stellar-to-total mass ratios based on weak lensing 
are now
beginning to emerge, and these can better inform our analysis, by allowing us to incorporate a total-to-stellar mass ratio that
is a function of stellar mass. 
Table \ref{pair_sum} summarizes the number of pairs selected by adopting a variable
total-to-stellar mass ratio using the weak-lensing based results presented by \nocite{leauthaud11} Leauthaud et al. (2011), and provides a comparison
with the other methods we have described.
We emphasize that weak-lensing measurements are constructed from large samples in the
appropriate mass range for our purposes, but at present such investigations are independent of galaxy morphology.
The results presented 
\citet{leauthaud11} are based on COSMOS imaging data \citep{scoville07} coupled to follow-up
spectroscopic redshifts from the zCOSMOS survey \citep{lilly07}. In Figure~\ref{sv-diagram_new},
we incorporate these data in attempt to improve upon Figure~\ref{sv-diagram_gon}.
The format of Figure~\ref{sv-diagram_new} is identical to that of Figure~\ref{sv-diagram_gon}, except that
non-constant stellar-to-total mass ratios have been used to infer the virial radii.
As a result, there are 31 physical pairs out of 148 pairs shown in Figure~\ref{sv-diagram_new} and 
again we flag merging pairs using filled symbols.
\citet{leauthaud11} probes systems with halo masses down to below $10^{11} M_{\odot}$, making this an excellent fit to our sample. 
However, it should be emphasized that weak lensing analyses are fairly new and subject to a large number of subtle systematics,
so some caution should be used in the interpretation of Figure~\ref{sv-diagram_new}. We suspect that
Figure~\ref{sv-diagram_new} is likely to have more long-term value than Figure~\ref{sv-diagram_gon}, but
the latter provides a robust backstop against which Figure~4 can be compared. 

\vspace{1cm}
\section{DISCUSSION}
\label{discussion}

\subsection{Abundance of physical pairs}

A comparison of Figures~\ref{sv-diagram_gon} and \ref{sv-diagram_new} 
makes it clear that the use of a non-constant mass-to-light ratio based on weak lensing data results in an increase
in the number of physical pairs. This is because the typical total-to-stellar mass ratios for galaxy-scale halos 
are much higher than the corresponding value for groups or clusters, so the derived virial radii
are also larger.
In Figure~\ref{sv-diagram_new}
one physical pair is a red-red system,
four physical pairs are blue-red systems, and 25 physical pairs are blue-blue systems.  
Although red-red pairs are clearly rare, their absence is not due to our experimental design: when we augmented
the IMACS data with Palomar observations 
purposely designed to find red-red systems with our technique, we did find them:
the Palomar observations resulted in five red-red pairs and one blue-blue pair, 
and the redshift success rate is consistent with that of IMACS observations\footnote{We again emphasize that Palomar observations are not included in the summary
statistics presented in this paper (e.g. in any of the discussion below) since they were intentionally biased 
and mainly useful for testing our methodology.}.

Overall, 30 out of 148 spectroscopic pair candidate pairs presented in Figure \ref{sv-diagram_new} turn out to be physical pairs. 
While the majority of these occupy the light grey region in the diagram which Patton~et al. (2000) suggest corresponds to systems with clear indications of tidal disturbance, 
around 10\% of the physical mergers lie outside this region.
These are typically pairs with one rather massive galaxy (resulting in a total system halo masses greater than $10^{12} M_{\odot}$,
denoted by the triangle symbols in Figure \ref{sv-diagram_new}), 
although two are pairs with relatively low velocity differences but larger separations.
The latter galaxies would likely be missed by the traditional close pair selection methods for identifying physical merging pairs, 
since it would be rather risky to select merging pairs by simply selecting on the basis of higher velocity differences or larger separations,
which would result in many false detections. 
It is also interesting to note that we find a number of systems ($\sim 15\%$) that are unlikely to be physical pairs even
though they are located in the bottom left corner of Figure~\ref{sv-diagram_new}. These 
are likely to be `false positives'
inferred using the Patton et al. (2000) methodology, and these systems emerge because their halo masses 
are low, resulting in small virial radii. Of course our method has false positives of its own:
since the velocities used in our analysis are line-of-sight,
and therefore a combination of projected galaxy peculiar velocities and cosmological redshifts, 
some fraction of these pairs are likely to have true separations of several Mpc and are unlikely to merge.
Such false positives are inevitable in any physical pair selection technique,
and the net result is that the number of close physical close pairs is likely to be
slightly over-estimated.

It is worthwhile to consider the ultimate fate of the physical pairs identified in the present survey. 
We assume that they will all eventually merge and become single systems. 
However, as noted in the Introduction, 
the timescale over which this will occur is far from certain, 
and the question of whether mergers trigger or suppress star-formation remains open. 
We will have more to say about this later in the discussion in the context of the Hot Halo model, 
but for now it is worthwhile to point out that Figure~\ref{sv-diagram_new} reveals a rather striking discrepancy in the fraction of red-red `pairs' that are likely to be line-of-sight optical superpositions, 
relative to the number of physical red-red pairs.
This is clearly seen even if one simply uses the straight-forward definition for a physical merger suggested by Patton et al. (2000),
and compares the fraction of pairs that are red-red systems inside the box ($\sim$ 1\%) to the corresponding fraction (5\%) of pairs that are red-red systems in regions of the diagram that are outside the box.

\subsection{Comparison with DEEP2 and VVDS Surveys}

Because the overall completeness of our survey (shown in Figure~\ref{dis_SF}) is relatively low,
we have mainly focused on making {\em relative} comparisons between the properties of physical pairs and field galaxies, 
since the photometric and spectroscopic selection criteria for the two populations are the same. 
However, it is worthwhile giving some consideration to the absolute numbers of objects seen, 
with the merger rate density of close pairs obtained from other surveys. Comparison with results from a merger
analysis of the DEEP2 survey \citep{lin08} is particularly
interesting, because these authors 
also selected close pairs based on dynamical criteria (i.e., separations $< 30 h^{-1}$ kpc and velocity difference $< 500$ km~s$^{-1}$)\footnote{Note that the maximum separation of close pairs selected in the DEEP2 survey is $10 h^{-1}$ kpc larger than the projected distance selection criterion used in this work.}.

The present IMACS survey's co-moving volume is two square degrees over a redshift range of $0.1 < z < 1.2$, which corresponds to $1.1\times10^7$ Mpc$^3$. 
Using the merger rate density of $\sim 1.0 \times 10^{-3} h^3$ Mpc$^3$ Gyr$^{-1}$ (where $h = 0.7$) and 
a merging timescale of 0.5 Gyr for all pairs taken from Figure~7 in \citet{lin08},
one expects  $\simeq 1900$ close pairs in our survey volume.
Figure~\ref{sv-diagram_new} shows that our survey contains 31 physical pairs. 
While this number seems low, 
in fact it matches our expectations based on the selection criteria
presented in Figures~\ref{dis_SF} and~\ref{SF}. 
In Figure \ref{SF}, near the peak of the galaxy count distribution at around $i\sim22.5$ mag the spectroscopic completeness is around $\sim 40\%$ for an individual galaxy, 
and this should be squared to compute the pairwise completeness (as a first 
order approximation, to account for the fact that we require successful redshifts for both galaxies in the pair).
So the overall spectroscopic completeness of galaxies targeted with slits is around 16\%. This must in 
turn be multiplied by the sparse sampling completeness factor which accounts for the fact that only
a subset of galaxies was targeted with a slit.
In Figure \ref{dis_SF} we see that the ratio of the parent sample to the observed sample is roughly 10:1 for potential pairs with 
separations smaller than $20 h^{-1}$kpc.
Therefore, our 31 physical pairs maps to 31/0.16$\times 10$ = 1937, which is in good agreement with the expected
$\simeq$ 1900 physical pairs 
predicted from the merger rate density of the DEEP2 survey.

Our main conclusion (that dry mergers are rare at intermediate redshifts)
has been foreshadowed by \citet{lin08}, using the data from the DEEP2 Survey.
These authors report that the red-red pair merger rate density within the redshift range 
$0.2 < z < 1.2$ is $\sim 1.0 \times 10^{-4}\, h^3\rm{Mpc}^{-3}\, \rm{Gyr}^{-1}$ with the merging timescale of 0.5 Gyr.
Using this red-red pair merger rate density and a merging timescale of 0.5 Gyr, 
we can estimate the number of red-red pairs predicted from the DEEP2 survey result given our IMACS survey volume.
Again using the IMACS survey volume noted earlier of $1.1\times10^7$ Mpc$^3$,
we would expect $(1.0 \times 10^{-4}\, h^3\rm{Mpc}^{-3} \rm{Gyr}^{-1} ) \times (0.5~\rm{Gyr}) \times (1.1 \times 10^7\, \rm{Mpc}^3) \sim 190$ red-red pairs
if our survey had placed a slit on every available galaxy and extracted redshifts with 100\% completeness. 
As we have already shown, the two completeness correction factors encapsulated by Figures~\ref{dis_SF} and~\ref{SF} can be used to
map this expectation into an observed number of pairs in our survey.
For red galaxies, our redshift completeness is 25\% at $i' \sim 22.5$ mag, 
and as noted before this should be squared to compute the pairwise completeness and this then must
be multiplied by the inverse of our sampling completeness.
Therefore our single detection of red-red close pair translates to $1/0.06\times 10 \simeq 160$ pairs after applying the completeness correction factors, 
which is consistent with the number of predicted red-red pairs obtained from the DEEP2 survey.

Some of our conclusions can also be tested using published results from 
the  VIMOS VLT Deep Survey (VVDS)  \citep{deravel09}.
These authors
provide a set of 36 spectroscopically confirmed physical pairs with separations r$_p < 20 h^{-1}$~kpc and  velocity 
differences $\Delta v < 500$km s$^{-1}$.
The fraction of pairs corresponding to early type galaxies 
is $\sim$8\%, which is significantly lower than the fraction  (25\%) 
estimated when the pair criteria are expanded to separations up to r$_{p}$ $=100 h^{-1}$kpc. 
Evidently, most of the red-red pairs in the sample are seen at
larger separations.

\subsection{Damp vs. Dry Mergers}

Based on the analysis presented in the previous
section, we conclude that there is a high degree of consistency in the predicted number of {\em early-stage} mergers (with their separations smaller than $20 h^{-1}$ kpc) seen in our pair analysis data, compared with the published space densities for such systems. It is less clear that
the {\em late-stage} mergers probed by morphological studies (e.g. that of Chou et al. 2011) are also consistent with
these numbers, given the very high-fraction ($>50\%$) of nearby late-stage mergers that are red. 

Why are so many nearby late-stage mergers red?
A possible solution might be the existence of a
short-lived star formation epoch 
during the early merging stage. In this scenario, 
some galaxies in the field are not truly `dry' nor truly `wet', since they
still contain a small amount of gas \citep{rampazzo05, crocker12}.
When two galaxies interact at the close pair stage, a small amount of star formation is triggered by tides, 
which makes at least one of the galaxies appear blue. This nascent star-formation
period must be fairly transient, since during the late-phase of the merger the remnant   
must become red in a short period of time (of order the merging timescale, $\sim 0.5$ Gyr).  The
result is a relatively high number density of dry mergers despite a
paucity of red-red pairs.  This begs the question: what combination of physics might result in 
weak tides that trigger star-formation, along with other processes that truncate it? One
possibility is a combination of tidally-induced merging coupled with 
the hot halo picture, as discussed in \S\ref{hothalo} below. However, before 
describing this idea further,
we must first consider the precise definition of a dry merger, both in our paper and
in the literature.

In this context, it is interesting to consider how the rest-frame $(g'-i')$ color selection threshold
we have used to define our red galaxy sample compares with the colour-selection criteria adopted in
other papers that also probe `red mergers' at intermediate redshifts.
Our $(g'-i')$ colour threshold has relatively
little sensitivity to contributions made by young stellar populations, so it is particularly interesting to
compare our work to \citet{lin08} [hereafter Lin08], who used a $(u-b)$ color threshold, and \citet{dokkum05} [hereafter vD05], who used
a $(b-r)$ color threshold, to define their red populations.
To inter-compare the surveys, we can use spectral synthesis models to investigate how much additional star-formation is needed in order for an underlying
old stellar population to no longer meet the red galaxy color-selection thresholds used in each of these papers.
We  used the BC03 code \citep{bc03} to construct  spectral energy distributions comprised of a dominant very old 
stellar population polluted by star-formation from a  
very young stellar population.
The stellar mass of the template galaxy was fixed to a stellar mass of $5 \times 10^{10}$ M$_{\odot}$.
Both stellar populations were modelled using exponential star-formation histories
with an e-folding timescale of 1~Gyr and a foreground-screen dust extinction model with $\tau_{V}=1.0$.
The ages for the old and the young stellar populations were fixed at 10 Gyr and 0.1 Gyr, respectively, and
we varied the stellar mass fraction between the old and the young populations before measuring the corresponding broad band colors.
Our analysis shows that the $(g'-i')$ color threshold used in the present paper is rather liberal --- 
a 10\% contribution to the total mass is needed from the young
stellar population in order to exceed our colour threshold.
In contrast to this, the 
$(u-b)$ color threshold used by Lin08 can accommodate a $\sim 1\%$ young stellar population, and
the $(b-r)$ color threshold used by vD05 can accommodate essentially no young stellar population ($<1\%$). 
This also implies that these early type galaxies may not be completely dry. 
While it is a mistake to assume that morphological characteristics uniquely map onto particular stellar populations, we suspect that
our red galaxy sample is likely to be comprised of a combination of elliptical galaxies and
early-type spirals,
while the Lin08 and vD05 samples are probably dominated by pure ellipticals. And if the
Lin08 and vD05 studies are of `dry mergers', our study is mainly made of systems better described as
`damp mergers'. 
However, it is worth noting that a non-negligible fraction of the red galaxies presented in vD05 
exhibit excesses in 8 and $24\mu$m bands \citep{desai11}, 
which suggests that at least some of these objects are star-forming. 
Putative early type galaxies may appear to be almost dry using rest-frame visible color selection criteria, 
but they may not be completely dry and/or may be undergoing gas rich minor mergers \citep{desai11}.
The present paper shows that even damp mergers are quite hard to find, so the truly dry mergers must be
remarkably rare.

\subsection{Implications for the Hot Halo Model}
\label{hothalo}
We conclude this section by noting the constraints placed on the
hot halo quenching model by our results.
As noted in the Introduction, in the hot halo quenching model 
systems with halo masses greater than $10^{12} M_{\odot}$ shock-heat 
infalling gas within the virial radius, so star-formation is quenched.
We have estimated the halo masses of all physical pairs,
and are able to subdivide our physical pair sample into two groups,
with  halo mass greater than and less than $10^{12} M_{\odot}$,
and denote these with different symbols in Figures \ref{sv-diagram_gon} and \ref{sv-diagram_new}.
These figures show no significant evidence for an enhancement in the fraction of red pairs with their projected separations smaller than $20 h^{-1}$ kpc in the high-mass group. 
Indeed, as noted earlier,
red-red pairs are simply rare at all masses, and there is an excess of blue systems, again at all masses,
relative to the field. 
Rather than being quenched, it seems that star formation is 
enhanced by galaxy-galaxy interaction at the close-pair (early stage merger) phase.  This begs the following question:
why is there so little evidence for shock-induced star-formation quenching in our data, given
the abundance of red late-stage mergers reported by Chou et al. (2011)? 
It is possible that the quantity of gas flooding into a potential well in a
merger is sufficiently large that shocks are inefficient until the merger
is nearly complete. The crossing
timescale of a close pair is only $\sim 60$ million years, and  
it may be 
difficult to increase the gas temperature in such a short time period by gravitational heating.
This admittedly rather vague suggestion does hint at the possibility that timescales
do matter, so that the degree of shock-induced quenching
might depend strongly on the phase of the merger being witnessed. 
Testing this intriguing idea will require a much larger sample 
than that used in the present paper.

Another possible explanation is that the gas in these interacting galaxies is not falling into the pair's dark matter halo;
instead, it is already present in the halo, and quite likely being accreted onto the central regions of one or both galaxies, 
as predicted by simulations \citep{mihos94,dimatteo07} and seen observationally \citep{barton00,ellison11}

\section{CONCLUSIONS}

We have obtained spectra for $\sim 2800$ candidate close pair galaxies at $0.1<z<1.2$
identified from the Canada-France-Hawaii Telescope Legacy Survey fields.  
Spectra of these systems were obtained using the multi-object spectrograph IMACS on the 6.5m Magellan 
and DBSP on the 5m Hale telescopes.
These data allow us to constrain the rate of dry mergers at intermediate redshifts and 
to test the `hot halo' model for quenching of star formation.
Redshifts were obtained for $\sim 50\%$ of the galaxies in our sample (1385 galaxies), and 
$\sim 80\%$ of the redshift measurement (1115 galaxies) have confidence levels greater than unity.
Because confirmation of physical pairs requires successful high-confidence redshifts for both
galaxies in the pair, in the end we are left with 148 close pair candidates (296 galaxies).
We used virial radii estimated from the correlation between dynamical and stellar masses
published by Leauthaud et al. (2011) as a reference to select physically merging
systems from this sample, based on halo occupation.
We find that around 1/5 of our candidate pairs (31 pairs) are physically associated
and share a common dark matter halo. 
These pairs are divided into red-red, blue-red and blue-blue systems
using rest-frame $g' - i'$ colors, using the classification method introduced in Chou et al. (2011).
After correcting for known selection effects, the 
fraction of blue-blue pairs is significantly greater than that of red-red and blue-red pairs.
Given a fairly liberal rest-frame color selection criteria in selecting red galaxies, 
red-red pairs are almost entirely absent from our sample, suggesting that red early-phase mergers are rare at $z\sim0.5$. 
This result is consistent with that obtained from the DEEP2 survey \citep{lin08}.
Our data supports models with a short merging timescale ($<0.5$ Gyr)
in which star-formation is enhanced in the early phase of mergers, but
quenched  in the late phase of mergers. Hot halo models may explain
this behaviour if virial shocks that heat gas are inefficient until the merger
is nearly complete. 
  
\bibliography{ms1}
\end{document}